\documentclass[preprint,12pt]{elsarticle}
\usepackage{geometry}
\usepackage{amssymb}
\usepackage{lipsum}
\usepackage{float}
\usepackage{subcaption}
\usepackage{lineno}    
\usepackage{setspace}  
\usepackage{natbib}
\setcitestyle{square, numbers}
\usepackage[colorlinks=true,linkcolor=blue,citecolor=blue]{hyperref}
\journal{Results in Physics}

\begin{document}

\begin{frontmatter}
\title{Design Optimization of Triple Gas Electron Multiplier for Superior Gain and Reduced Ion Backflow}
\author[inst1]{Sachin Rana\corref{cor1}}
\ead{sachin.rana@cern.ch}
\author[inst1]{Md. Kaosor Ali Mondal}
\ead{mohommad.mondal@cern.ch}
\author[inst1]{Poojan Angiras}
\ead{poojan.angiras@cern.ch}
\author[inst1]{Amal Sarkar}
\ead{amal.sarkar@cern.ch}
\cortext[cor1]{Sachin Rana}
\affiliation[inst1]{organization={Indian Institute of Technology, Mandi},
            addressline={}, 
            city={Kamand},
            postcode={175005}, 
            state={Himachal Pradesh},
            country={India}}
\begin{abstract}
Micro-Pattern Gas Detectors (MPGDs) are widely employed in modern high-energy and nuclear Physics experiments due to their excellent spatial resolution, high rate capability, and operational stability. Among these, the Gas Electron Multiplier (GEM) has emerged as one of the most widely adopted MPGD technologies. Despite their widespread adoption, GEM detectors based on the conventional bi-conical hole geometry do not always achieve optimal performance, particularly in maximizing effective gain while suppressing ion backflow. One of the primary factors limiting a GEM's performance is ion backflow. The accumulation and gradual discharge of these ions might alter the local electric field, resulting in a temporary dead time and complicating responses to subsequent events. These limitations pose challenges for applications that require high precision and stable, long-term operation. In this work, we address these issues by investigating modified GEM geometries designed to enhance gain performance and reduce ion backflow, thereby improving overall detector performance. The current study investigates geometric optimization strategies for a triple-GEM detector to enhance performance, mitigate ion backflow, and augment gain. The detector structures were designed using the ANSYS Mechanical APDL, and the associated electrostatic field configurations were computed using the ANSYS Maxwell. A thorough investigation of gain and ion backflow calculations was carried out when the generated field maps were interfaced with Garfield$^{++}$. The potential enhancements in detector efficiency and stability offered by the proposed modifications to the GEM foil geometry provide valuable insights for the design of next-generation gaseous detectors.
\end{abstract}




\end{frontmatter}




\section{Introduction}
\label{introduction}

The development of sophisticated detector technology has reached previously unheard-of levels of complexity and precision, thanks to the physics processes investigated in current and upcoming high-energy and particle physics investigations. These investigations place a significant strain on detection systems due to their high collision rates and extraordinary event multiplicities \cite{grupen2008particle,tsyganov2008gas} in experimental nuclear and particle physics. Detectors must have outstanding momentum and jet energy resolution, robustness under extreme radiation exposure \cite{bressan1999beam}, exceptional spatial and temporal resolution, and reliable particle identification capabilities to reliably reconstruct and identify the enormous number of particles generated in such environments. Additionally, they must ensure that the system operates consistently for extended periods while simultaneously reducing the impact of background noise and ion feedback that could compromise data integrity.

Gaseous detectors, particularly Time Projection Chambers (TPCs) \cite{marx1978time,hilke2010time}, have been a key component of various detector technologies developed to address these issues, thanks to their ability to provide three-dimensional tracking with high precision and low material budget. Numerous large-scale experiments have relied heavily on the TPC concept, which enables the reconstruction of charged-particle paths by drifting ionization electrons toward a readout plane under a uniform electric field. However, the performance requirements for gaseous detectors have increased significantly as experimental settings shift toward higher luminosities and event rates, necessitating advances in their readout and amplification methods. Due to the phenomenon of ion backflow (IBF), traditional TPCs that use Multi-Wire Proportional Chambers (MWPCs) \cite{charpak1970some,sauli1977principles,charpak1979multiwire} for gas amplification have major difficulties at high rates. In MWPCs, the positive ions produced in the amplification region migrate back into the drift volume during the avalanche process, altering the local electric field and influencing the homogeneity of the electron drift paths. Gating grids can reduce this effect, but they also introduce dead time by nature, which reduces the detector's rate capabilities. Over the past few decades, significant research and development efforts have been devoted to the advancement and application of Micro-Pattern Gaseous Detectors (MPGDs)~\cite{titov2013micro,shekhtman2002micro}. These detectors represent a category of sophisticated gaseous radiation detectors \cite{buzulutskov2007radiation} that utilize precisely engineered electrode geometries. These structures are fabricated using microelectronic and photo-lithographic processes, to attain superior spatial resolution, exceptional rate performance, and improved operational stability. Due to their inherent ability to confine gas amplification to microscopic regions and efficiently mitigate ion backflow, MPGDs have become a compelling alternative to traditional wire-based gaseous detectors. Consequently, substantial research and development efforts have been dedicated to employing Micro-Pattern Gaseous Detectors as readout and amplification elements in next-generation tracking systems, aiming to address the constraints of conventional technologies and facilitate uninterrupted, dead-time-free operation.

Introduced by F. Sauli at CERN in 1997 \cite{sauli1997gem,sauli2004progress,benlloch1998development,sauli2003development}, the Gas Electron Multiplier (GEM) is one of the most popular and successful MPGD devices. A GEM is made up of a thin, usually 50 µm thick polyimide (Kapton) foil that is coated with copper on both sides and has a dense array of tiny holes in a regular hexagonal pattern. A powerful electric field, on the order of 50–100 kV/cm, forms inside the holes when a potential difference of approximately 400–500 V is applied between the two copper layers. Charge amplification occurs when the electric field accelerates the electrons produced by the initial ionization of the gas, leading to an avalanche multiplication process \cite{bhattacharya20173d}. The effective gain of a single GEM foil typically ranges from 20 to 100, depending on the operating voltage, hole shape, and gas combination. Double or triple-GEM configurations \cite{fallavollita2019triple,bonivento2002complete} can be created by cascading numerous GEM foils in series to maximize total gains while maintaining stability and reducing the likelihood of discharge \cite{bachmann2002discharge,utrobicic2019studies}. A drift electrode, three GEM foils, and a segmented anode readout plane make up the triple-GEM architecture \cite{bressan1999high}, which offers a total gain of between $10^{4}$ and $10^{5}$. These detectors have a detection effectiveness of more than 98\% with outstanding temporal and spatial resolution \cite{kudryavtsev2017study,kudryavtsev2020spatial}. Furthermore, GEM-based detectors exhibit high-rate capability, cost-effectiveness, and good radiation tolerance, making them ideal for use in large-area systems in a variety of experiments \cite{sauli2016gas} at CERN, BNL, FAIR, and potential future colliders.



Although GEM-based detectors have demonstrated their effectiveness \cite{kumar2024study}, additional optimization is necessary to achieve consistent gain, sustained operational stability, and efficient suppression of ion backflow, especially in high-rate \cite{bachmann2001optimisation} and large-area applications. Ion backflow is a significant performance-limiting phenomenon in GEM detectors, occurring when a portion of the positive ions generated during the avalanche multiplication in the high-field regions of the GEM cavities migrates back toward the drift or conversion region. The buildup of these back-drifting ions generates space-charge effects that locally distort the electric field, thereby impairing electron transport, reducing spatial and energy resolution, and causing temporal anomalies in the detector response during continuous irradiation. Consequently, unregulated ion backflow \cite{lyashenko2006advances,lyashenko2007further} significantly limits the attainable rate capability and reliability of GEM-based systems. A key factor in the detector's performance in this context is the geometric arrangement of the GEM foil, which is specified by characteristics such as hole pitch, foil thickness, and aperture shape. These geometric features determine the local electric field distribution within and surrounding the GEM holes, which subsequently influences electron collection efficiency, transmission, avalanche gain, and the extraction and transport of positive ions. Consequently, even minor adjustments to the geometry of the GEM foil can produce notable enhancements in suppressing ion backflow  \cite{maia2004avalanche}, gain performance, and overall detector efficacy. As a preliminary investigation, a comprehensive analysis was performed on a single-GEM detector, in which the foil hole geometry was modified from the conventional bi-conical configuration to a solitary conical design~\cite{Sarkar:2025tvu,mondal2025optimizing}. The thickness of the bottom copper layer was systematically altered, and the ratio of ion backflow to effective gain was examined. The performance of the modified single-GEM configuration was then compared to that of a standard single-GEM detector including bi-conical holes. The findings demonstrated a significant enhancement in effective gain and a substantial reduction in ion backflow, indicating that the suggested geometric modifications are superior.

Based on the encouraging results of the single-GEM study \cite{Angiras:2025fcr}, this work extends the same design methodology to the development of a triple-GEM detector. The traditional bi-conical hole shape has been replaced with a single conical structure, and the thickness of the lower copper layer has progressively increased from a typical 5 $\mu$m to 10 $\mu$m and then to 15 $\mu$m. These modifications alter the symmetry and intensity of the electric field inside the amplification channels, thereby influencing electron multiplication and ion mobility. The structural modifications were implemented in a conventional triple-GEM configuration. The impact on the detector's operating properties was methodically examined and contrasted with that of a conventional triple-GEM design. The electrostatic and charge transport aspects of these modified GEM foils were investigated through an extensive computational campaign. First, comprehensive electric field maps inside and around the GEM holes for various geometric configurations were calculated using finite element models \cite{huebner2001finite}. Garfield$^{++}$ \cite{Garfieldppusemanual}, an advanced package for modeling gaseous detectors at the tiny level, was then used to import the field maps. Garfield$^{++}$ simulates the creation of induced signals at the readout electrodes, as well as the drift, diffusion, and multiplication of electrons and ions in the detector gas under realistic electric field conditions. Using this methodology, a systematic analysis of the effects of variations in copper thickness and hole shape on key performance metrics, including ion backflow fraction, effective gain, and electron transmission, has been conducted in a triple-GEM arrangement. The findings demonstrate that the geometric arrangement of the GEM foils significantly affects the detector's amplification properties. The effectiveness of electron collection and multiplication is increased when the bi-conical hole geometry is changed to a single-conical hole. Furthermore, the percentage of ions that escape into the drift region decreases with increasing lower copper thickness. When considered collectively, these factors enhance ion backflow suppression and increase potential gains without compromising detector stability. The suggested changes demonstrate that minor structural improvements can produce significant performance gains, as confirmed by thorough Garfield$^{++}$ calculations. These enhancements are especially pertinent to next-generation high-rate experiments, where stable and accurate operation depends on concurrently retaining high gain and low ion feedback. This study provides a clear and impactful pathway for advancing GEM foil design in next-generation high-performance triple-GEM detectors.


The results presented in the subsequent sections demonstrate that geometrical optimization constitutes an effective strategy for enhancing GEM detector performance to meet the requirements of future experiments. Section 2 describes the fundamental operational principles of the Triple-GEM detector. The computational tools and framework used for this study are described in Section 3 and Section 4 provides a comprehensive and systematic comparison of the ion backflow-to-gain ratio between the traditional Triple-GEM and geometrically modified GEM configurations. Furthermore, a quantitative and objective comparison of ion absorption percentages is conducted to assess the impact of geometrical modifications on detector performance. Finally, Section 5 summarizes the principal findings of this study.

\section{Working Principle of Triple GEM}
Within a triple Gas Electron Multiplier (GEM) detector, the incident charged particle initially passes through the 3 mm drift region, where it ionizes gas molecules and generates primary electron–ion pairs. 
  \begin{figure}[H]
  \centering
  \includegraphics[width=0.65\columnwidth]{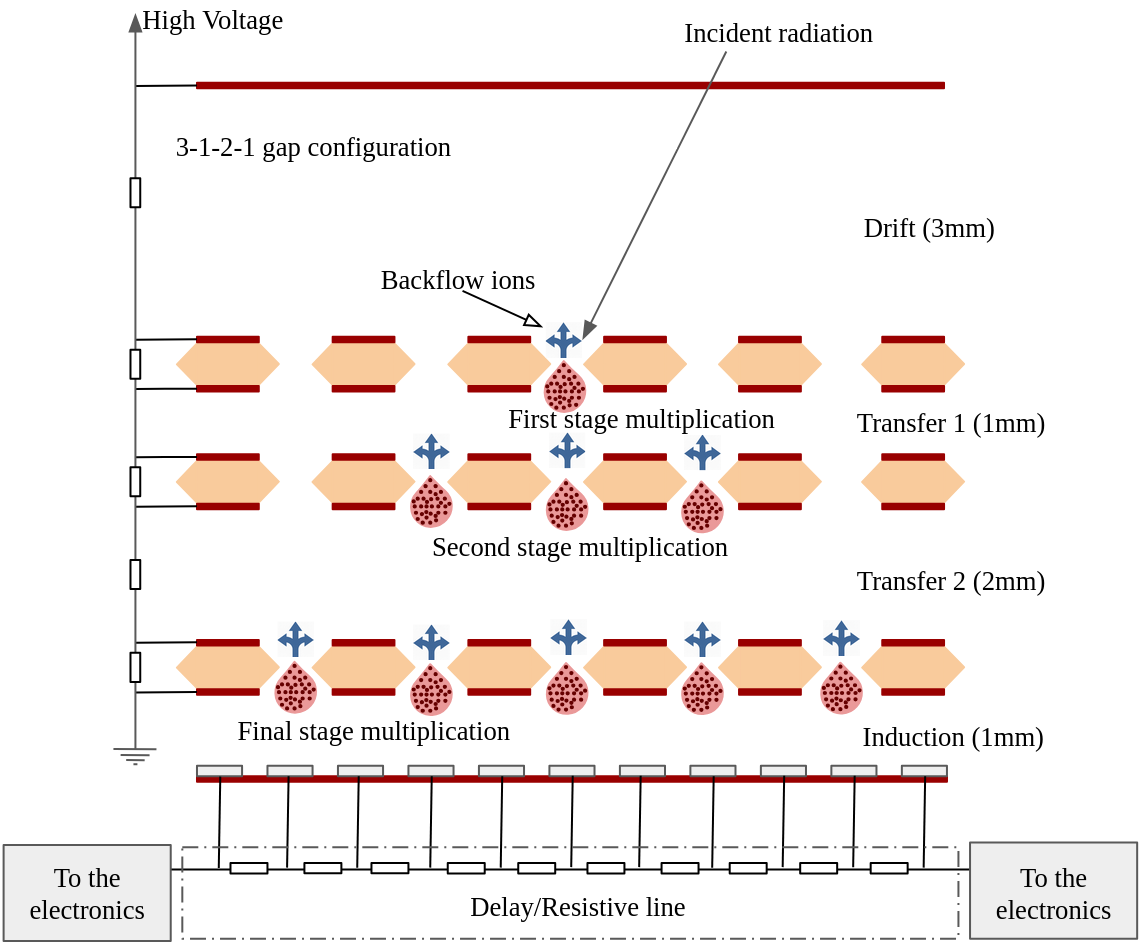}
  \caption{Schematic diagram of Triple GEM, showing the three 
stage multiplication. Multiplied electrons move towards the next foil and , then towards the induction region, and are finally to the electronics. Back-flowing ions get collected at respective copper layers, and a few move towards the Drift region, which contributes to the back current.}
  \label{fig:tri_gem}
\end{figure}
These pairs are directed toward the initial GEM foil under a relatively low electric field, facilitating efficient collection without inducing early multiplication. When electrons get to the first GEM foil, they go into the microscopic holes' high-field areas, where strong field gradients start the first stage of avalanche multiplication. Then, the cloud of amplified electrons comes out into the first transfer region, which is 1 mm wide, and is efficiently guided toward the second GEM foil by an optimized transfer field. At the second and third GEM foils, a similar process takes place: in each microscopic-hole structure, electrons undergo additional avalanche multiplication under fields of typically 50-60 kV/cm. This creates a multi-stage gain mechanism that increases the total electron yield and simultaneously reduces ion backflow at each intermediate GEM electrode through effective ion trapping.

Before electrons are carried via the 1 mm induction zone toward the readout anode, they undergo the last multiplication step in the third foil, where the transfer field in the second transfer region (2 mm) is fine-tuned to achieve high extraction and focusing efficiency. In this last gap, an induction field that is precisely chosen allows for fast electron collection to the readout plane while limiting additional ionization. In this way, when the multiplied electron swarm arrives, it induces a measurable current pulse, and most of the ions produced in each stage drift back and are captured by the electrodes of the corresponding GEM foils, thus reducing space-charge effects and preserving signal quality.
High gain, minimal ion backflow, and excellent spatial resolution are characteristics of triple-GEM detectors, which result from consecutive avalanche stages, effective inter-foil transfer, and regulated electron extraction. The complete process is shown in Fig.~\ref{fig:tri_gem}
  
\section{Use of Various Computational Tools:}
The geometrical characteristics of the foils significantly affect the gain and efficiency of a triple GEM detector.  Each GEM foil in the standard triple GEM configuration is consists of a 50 $\mu$m-thick Kapton layer clad with 5 µm of  both surfaces.  The holes are arranged in a regular hexagonal pattern with a pitch of 140 $\mu$m, i.e., the center-to-center distance between adjacent holes.  The bi-conical structure is characterized by each cavity having an outer diameter of 70 µm and an interior diameter of 50 µm.

All geometrical constructions of the GEM foils and detector configuration were implemented using ANSYS, while the subsequent analysis of effective gain and ion backflow was performed using the Garfield$^{++}$ framework. The methodology and results are discussed in detail in the following section.
\subsection{Geometry Construction using ANSYS Maxwell:}
ANSYS Maxwell \cite{martyanov2014ansys}, is a finite element method (FEM) based simulation tool, that was implemented to model and analyze the GEM geometry. Materials were assigned to each component, and appropriate boundary conditions were applied to construct the complete GEM detector geometry in ANSYS.

\begin{figure}[H]
  \centering
  \includegraphics[width=0.8\columnwidth]{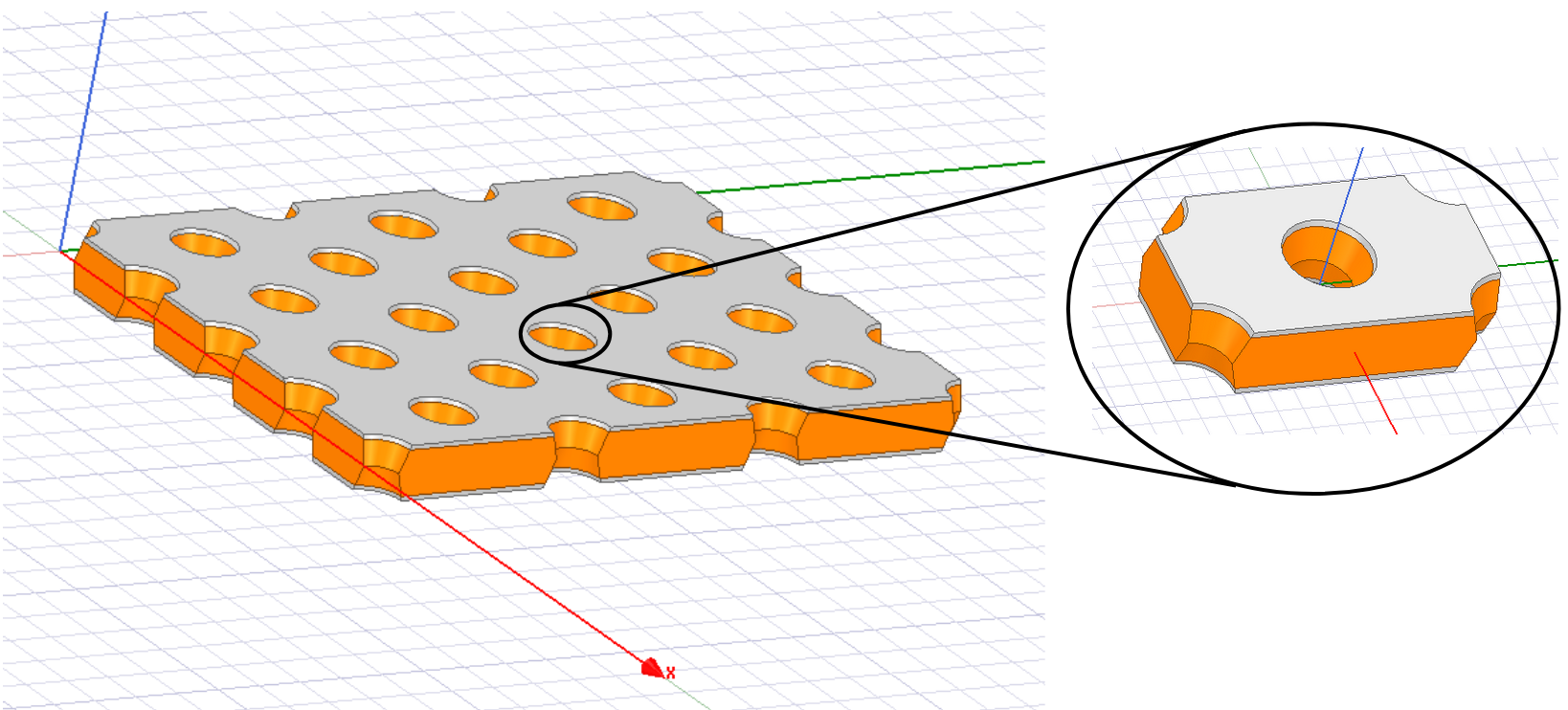}
  \caption{GEM foil geometry modeled using ANSYS Maxwell. The inset shows a single unit cell, whose periodic replication forms the complete foil structure. Stacking multiple such foils at specified inter-foil separations results in the complete triple-GEM configuration.}
  \label{fig:unitCell}
\end{figure}

The electric field distribution within the system was computed by setting voltages on the cathode, the upper and lower copper layers of all three GEM foils, and the anode plane. The drift gap, GEM holes, two transfer regions, and induction gap, which are responsible for primary ionization, electron multiplication, and signal collection, were used to determine the average electric field intensity in the detector.
Initially, a unit cell of the GEM foil was modeled with precise dimensions to ensure computational efficiency and accuracy. Then, by replicating the unit cell in both directions x and y, a complete foil structure is obtained.
The unit cell shown in Fig.~\ref{fig:unitCell} was replicated and merged to produce a hexagonal lattice structure that resembles the GEM foil of the desired dimension, following the definition of material properties. Subsequently, three foils were inserted into a simulated gas chamber filled with a standard working mixture of Ar:CO$_2$ (70:30). This mixture was selected because of its well-established equilibrium, characterized by low diffusion, high gain, and stability.

{The electric field profiles in different regions of the detector were thoroughly analyzed as a consequence of the applied potentials on the copper electrodes and the boundaries of the gas chamber.} Subsequently, the ANSYS field calculator was used to compute the field distributions, which served as critical input for subsequent investigations, including electron drift, avalanche development, and gain estimation.

\subsection{Generating the field maps using Mechanical APDL:}
ANSYS Mechanical APDL \cite{thompson2017ansys} is a scripted finite element analysis environment that enables the precise and flexible definition of complex geometries, materials, and boundary conditions. Its parametric and command-driven framework is particularly well-suited for detailed detector modeling and the generation of high-fidelity field maps for subsequent studies.
Within the APDL framework, three GEM unit cells were constructed with precisely defined separations corresponding to the first, second, and third GEM foils as shown in Fig.~\ref{fig:apdl}. The entire detector geometry was generated using APDL scripting, providing a flexible and efficient approach for parametric modeling. After defining all remaining regions and applying the appropriate boundary conditions, four output files were produced: ELIST.lis, containing information on the geometric elements forming the detector structure; MPLIST.lis, detailing the assigned material properties; NLIST.lis, listing the nodal information; and PRNSOL.lis, providing the electric potential or weighting field values at the nodes. 

These field map files collectively encapsulate the complete detector description and serve as essential inputs for subsequent studies, including gain and ion backflow studies, performed using Garfield$^{++}$.
\begin{figure}[H]
    \centering
    \begin{subfigure}[t]{0.48\columnwidth}
        \centering
        \includegraphics[width=\linewidth]{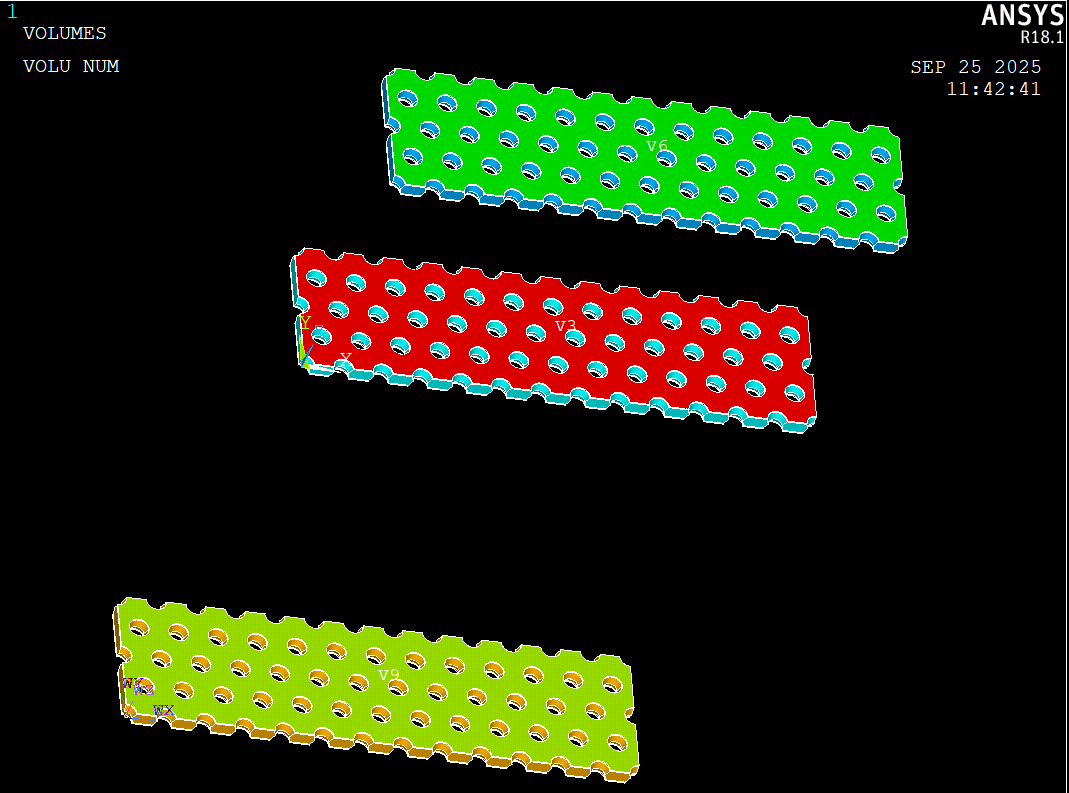}
        \caption{GEM foils modeled using ANSYS Mechanical APDL were positioned at predefined locations within the detector geometry}
        \label{fig:apdl}
    \end{subfigure}
    \hfill
    \begin{subfigure}[t]{0.48\columnwidth}
        \centering
        \includegraphics[width=\linewidth]{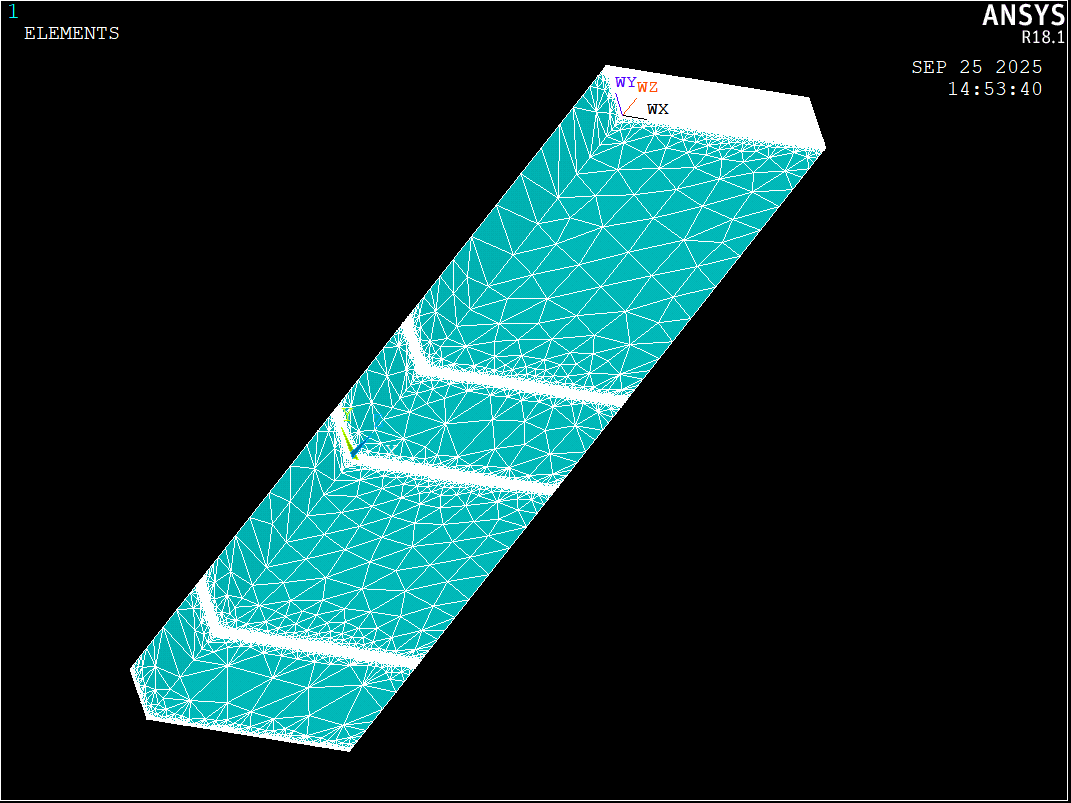}
        \caption{A meshing procedure was performed to discretize the complete detector structure and store detailed geometric information.}
        \label{fig:fullfoil}
    \end{subfigure}
    \caption{Three-dimensional detector geometry constructed using Mechanical APDL and employed for field map calculations.}
    \label{fig:apdl}
\end{figure}

\subsection{Connecting field maps to Garfield$^{++}$:}
The generated files in ANSYS Mechanical APDL (.lis files) are used in Garfield$^{++}$ for further study.
It excels at simulating the complex behavior of electrons and ions within the gas, accounting for the effects of electric fields, as well as the dynamics of charge multiplication and collection.

\begin{figure}[H]
  \centering
  \includegraphics[width=0.8\columnwidth]{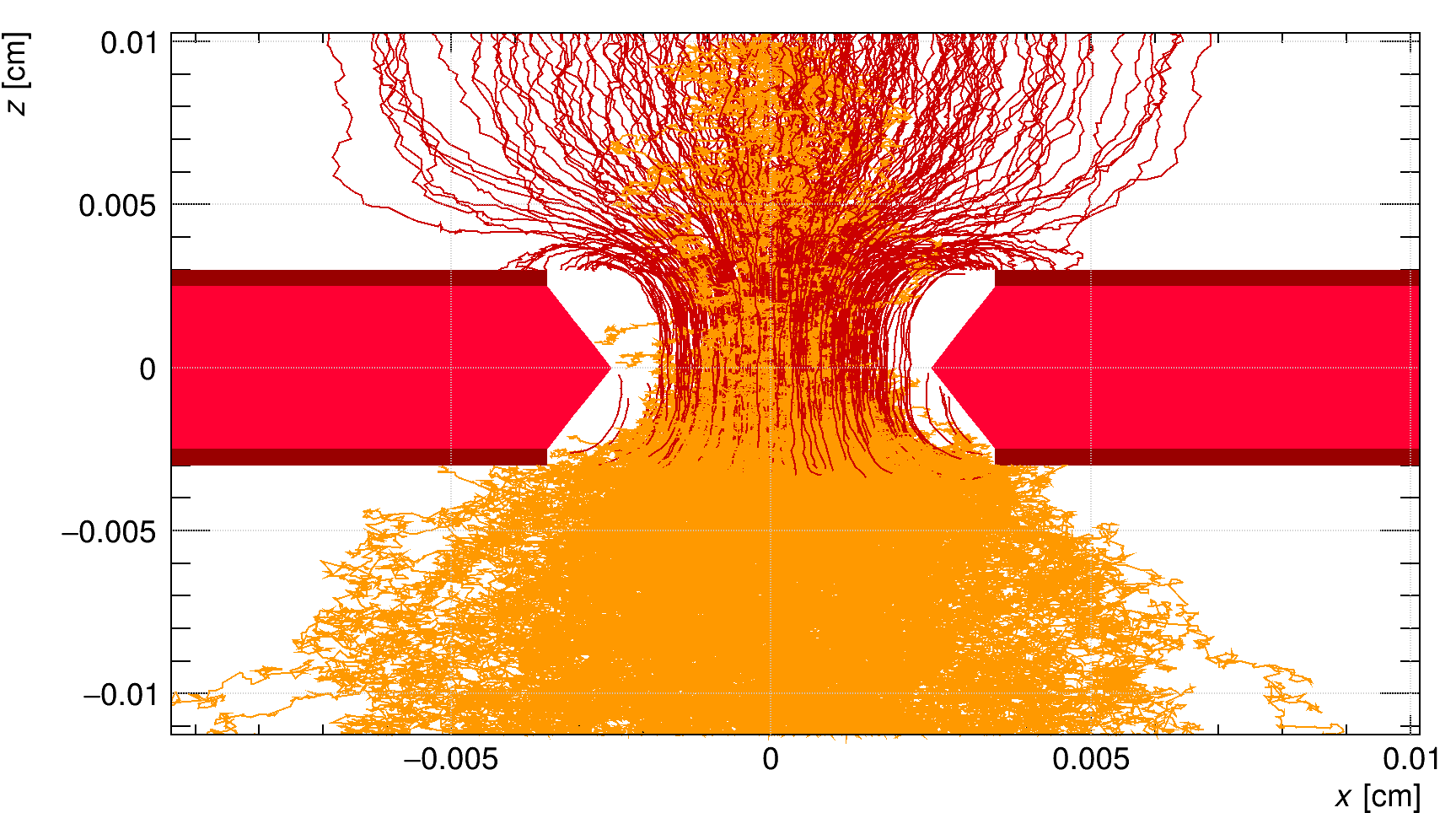}
  \caption{Visualization of the avalanche development within a single GEM hole simulated using Garfield$^{++}$. Electron trajectories are shown in yellow, while ion trajectories are shown in red. The electrons drift toward the readout electrode, whereas the majority of ions are absorbed at the copper layer, with a small fraction escaping back into the drift region}
  \label{fig:garfield}
\end{figure}

By utilizing advanced computational methods and seamlessly integrating with tools such as ANSYS and Magboltz \cite{al2020electron}, Garfield$^{++}$ provides a comprehensive and accurate representation of detector performance. It precisely models the electron avalanche process by solving the equations of motion for electrons and ions, accounting for the electric field, gas composition, and the detector's geometry. The \emph{.lis} (\emph{ELIST.lis, NLIST.lis, MPLIST.lis, PRSNOL.lis}) files are imported in Garfield$^{++}$ to study the behavior of ions and electrons. The study encompasses stochastic processes, including electron collisions, energy loss, and secondary ionization events, thereby providing a detailed and accurate representation of the avalanche phenomenon. 
This enabled us to predict gas detector performance, optimize their design, and gain a deeper understanding of the underlying physical processes. By numerically solving the Boltzmann transport equation, which models the microscopic motion of electrons undergoing successive collisions in a gas, the Magboltz program computes the transport properties of drift gasses. Under the high electric field within the GEM holes, electron multiplication occurs which can be seen in Fig.~\ref{fig:garfield}, leading to an avalanche that emerges from the foil's lower side. It can calculate the drift velocity, starting point, and end point of ions and electrons under the influence of electric and magnetic fields by tracking the virtual propagation distance of electrons. 
The yellow and red lines denote the drifts of electrons and ions, respectively. An electron with a lower ionization energy of e$_{0}$=1 eV is used to study GEM in Garfield$^{++}$. The electron's initial position is selected randomly. After the avalanche, the endpoints of the electrons and ions are calculated. If the endpoints of the ions are far above the upper copper foil, these ions are considered to be the backflow of the ions. The number of electrons produced in the avalanche process for a single-line spectrum is referred as gain. 

\section{Experimental verification}
A comprehensive validation investigation has been conducted to verify the dependability and prediction accuracy of the framework used in this work, by replicating the experimentally recorded gain characteristics of a conventional single Gas Electron Multiplier (GEM) detector setup.
\subsection{Detector Configuration}
The reference detector configuration \cite{bachmann1999charge} aligns with a commonly used conventional single GEM foil arrangement. The geometric specifications consist of a hole pitch of 140 $\mu$m, an inner hole diameter of 50 $\mu$m, and an outside hole diameter of 70 $\mu$m. The GEM foil has a 50 $\mu$m thick Kapton sheet interposed between the conductive copper electrodes. These settings align with standard experimental implementations documented in the literature.
 A uniform electric field is supplied in the drift area to direct primary electrons toward the GEM foil, while a robust electric field inside the GEM apertures, created by the voltage differential ($\Delta$V) between the GEM electrodes, facilitates electron multiplication via avalanche mechanisms. 
The experimental analysis indicates that the drift and induction fields are sustained at 1.6 kV and 5 kV, respectively, while the voltage difference across the GEM foil ($\Delta$V) is consistently altered. The gain is then assessed for various $\Delta$V values, maintaining constant drift and induction fields, to provide a direct comparison with experimental observations. The detector response is studied with an X-ray source of energy at 5.9 keV, often linked to Fe-55 radioactive sources used in experimental gain assessments. The interaction of X-rays with the gaseous medium creates primary ionization clusters, resulting in a certain quantity of primary electrons. The primary electrons then migrate toward the GEM holes under the influence of the applied electric field.
The detector shape and electrostatic setup are first established in ANSYS Mechanical APDL, where the electric potential and field distributions are calculated with great accuracy. The resultant field maps are produced as .lis files and then loaded into Garfield$^{++}$, used for precise microscopic monitoring of electron transit and avalanche phenomena. All geometrical dimensions, applied voltages, gas composition, and operating conditions were chosen to reproduce the experimental configuration reported in Ref.~\cite{bachmann1999charge} as closely as possible.
\subsection{Comparison with experimental data}
The effective gain of the detector is defined as the ratio of the total number of electrons produced in an event to the number of primary electrons produced by the original ionizing event.
\begin{figure}[H]
  \centering
  \includegraphics[width=0.65\columnwidth]{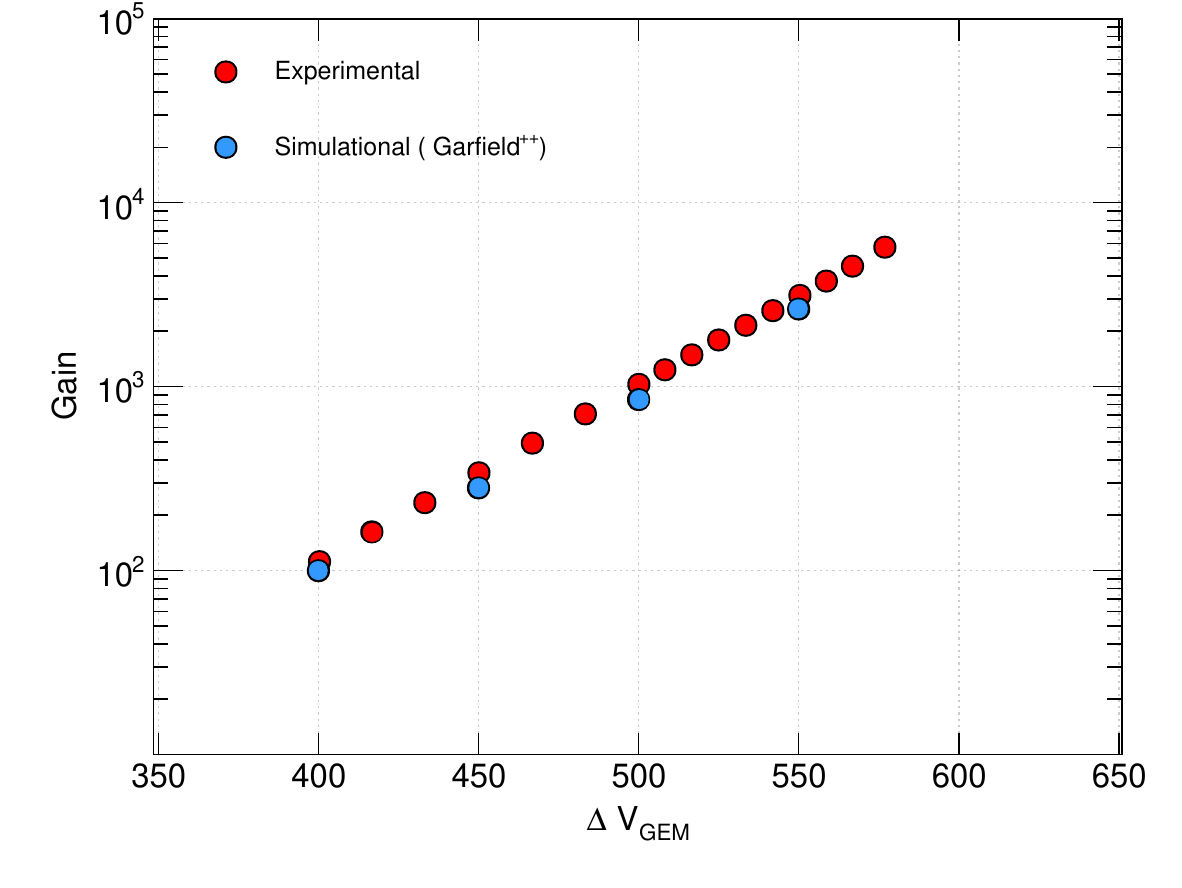}
  \caption{Gain as a function of the voltage difference ($\Delta$V) across a single GEM foil, comparing experimental measurements with simulation results obtained under identical conditions. The gain increases with $\Delta$V due to enhanced electron multiplication, and a good agreement between the two is observed, validating the computational model. Red dots represent the experimental data whereas blue dots represent the simulation results. }
  \label{fig:exp}
\end{figure}
The findings demonstrate a significant reliance on $\Delta$V, indicating the exponential characteristics of electron multiplication under high electric fields inside the GEM holes. A direct comparison of the computed gain values with the experimental data as shown in Fig.~\ref{fig:exp} demonstrates strong agreement over the examined voltage range. This consistency demonstrates that the framework used throughout the study effectively captures the fundamental physical mechanisms governing electron transit and amplification in GEM detectors. Minor discrepancies, if existent, may be ascribed to flaws in the real detector geometry, and intrinsic simplifications of the simulation model.
Nonetheless, the comprehensive agreement confirms the applicability of the current framework for further investigations and enhancements of GEM-based detection systems. The effective replication of experimentally obtained gain characteristics illustrates the reliability and precision of the integrated ANSYS–Garfield$^{++}$ method. This validation instills confidence in the framework's applicability for predictive analyses, including detector tuning, parameter exploration, and performance assessments across diverse operating scenarios.
\section{Results}
To optimize the geometry of the GEM foil and enhance the efficacy of the triple GEM detector, numerous investigations have been conducted. In the current investigation, extensive simulations were implemented to evaluate the influence of the hole shape on the detector's overall performance and gain. To optimize gain, a modified geometry was implemented. The copper thickness on the lower surface of the GEM foils was increased to enhance ion collection efficiency and minimize ion backflow, enabling additional studies. Lastly, a comparative analysis was conducted to assess the ratio of ion backflow to gain for the newly proposed geometries in comparison to the standard triple GEM configuration.
\subsection{Standard Triple GEM:
}
This research first assessed the performance of a typical triple GEM detector using a bi-conical hole shape.  The effective gain was determined from the number of simulated occurrences for five different initial electron energies.  
\begin{figure}[H]
  \centering
  \begin{subfigure}[b]{0.49\textwidth}
    \centering
    \includegraphics[width=\textwidth]{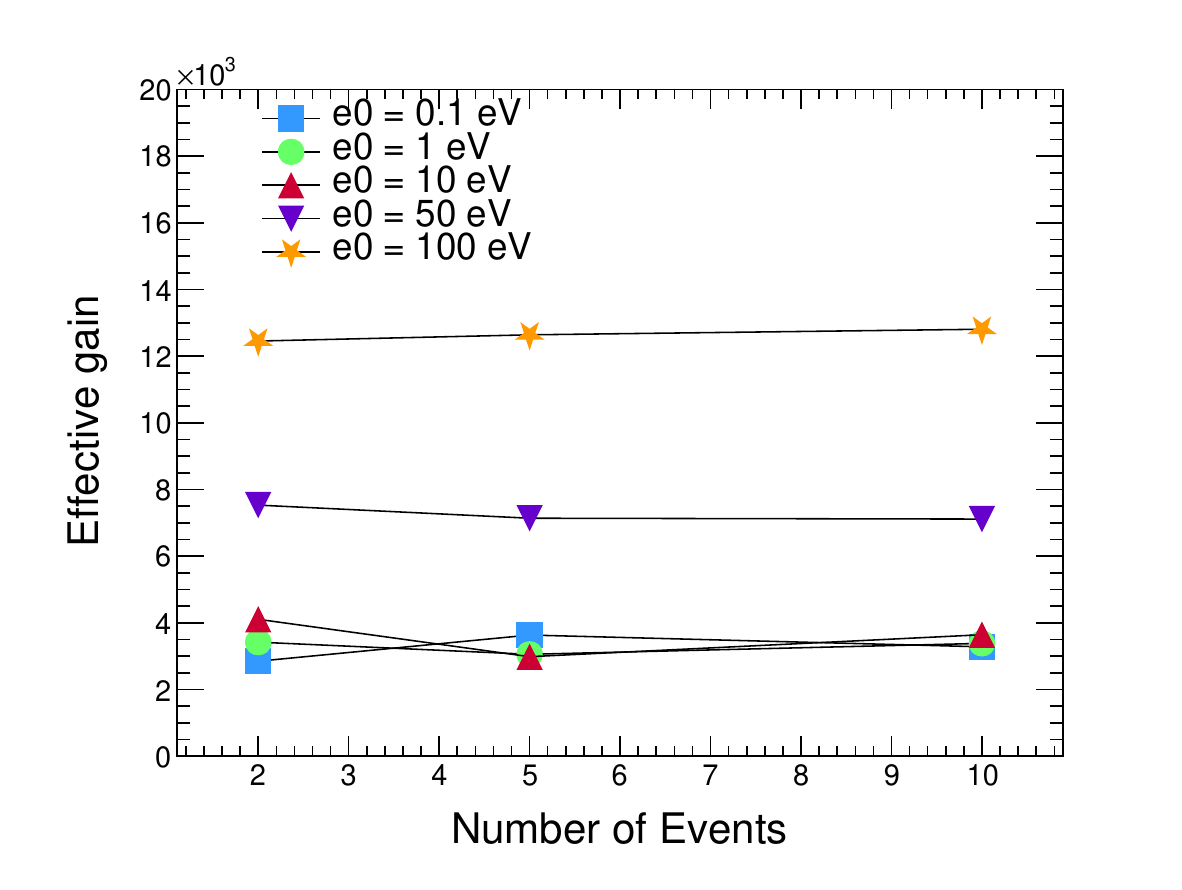}
    \caption{Effective gain of a standard Triple GEM evaluated for five different initial electron energies. For energies below the minimum ionization potential of the detector gas, Argon, the effective gain remains approximately constant. Once this threshold is exceeded, the effective gain increases with the initial electron energy.}
    \label{fig:eff_bc}
  \end{subfigure}
  \hfill
  \begin{subfigure}[b]{0.49\textwidth}
    \centering
    \includegraphics[width=\textwidth]{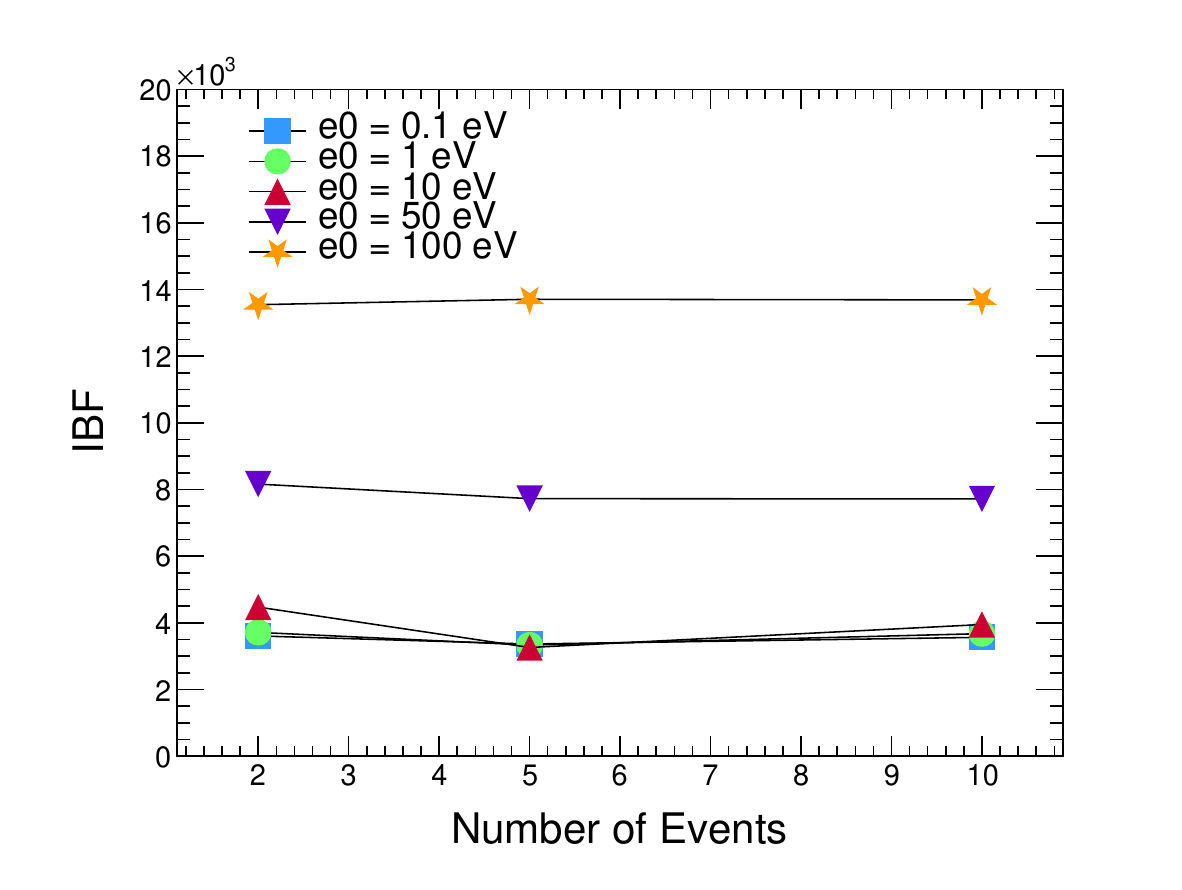}
    \caption{Ion backflow in a standard triple-GEM detector as a function of the initial electron energy, evaluated for five different energies. Below the detector gas's minimum ionization potential (argon), the ion backflow shows little variation. Above this threshold, a clear increase in ion backflow is observed with increasing initial electron energy.}
    \label{fig:IBF_a}
  \end{subfigure}
  \caption{Standard Triple GEM detector performance.}
  \label{fig:std_combined}
\end{figure}
The effective gain was found to be independent of the number of events across all examined energies, as shown in Fig.~\ref{fig:eff_bc} . This behavior validates the detector's statistical stability and repeatability under consistent operating conditions.
For the working gas argon, the effective gain demonstrated a practically constant value for beginning electron energies within the ionization threshold of the gas.  In this energy domain, primary ionization is constrained, and avalanche multiplication is primarily determined by the configuration of the applied electric field rather than the initial energy of the electrons.  Once the electron energy is above the ionization threshold, a significant increase in effective gain is noticed, as shown in Fig.~\ref{fig:eff_bc}.  This improvement may be ascribed to the initiation of supplementary ionization processes and enhanced avalanche formation inside the GEM holes at elevated energy. The ion backflow per event was assessed for the same standard triple GEM setup and beginning electron energies. Analogous to the effective gain, the number of ions migrating back into the drift zone was found to be independent of the total number of events, indicating consistent ion transport across the simulated dataset, as shown in Fig.~\ref{fig:IBF_a}.

For electron energies under the ionization threshold, the ion backflow remained very stable, indicating the restricted generation of secondary ions during avalanche formation. A substantial increase in ion backflow was observed at elevated electron energies. This increase is directly associated with larger avalanche sizes and the increased ion production resulting from higher-energy ionization processes. 
  \begin{figure}[H]
  \centering
  \includegraphics[width=0.75\columnwidth]{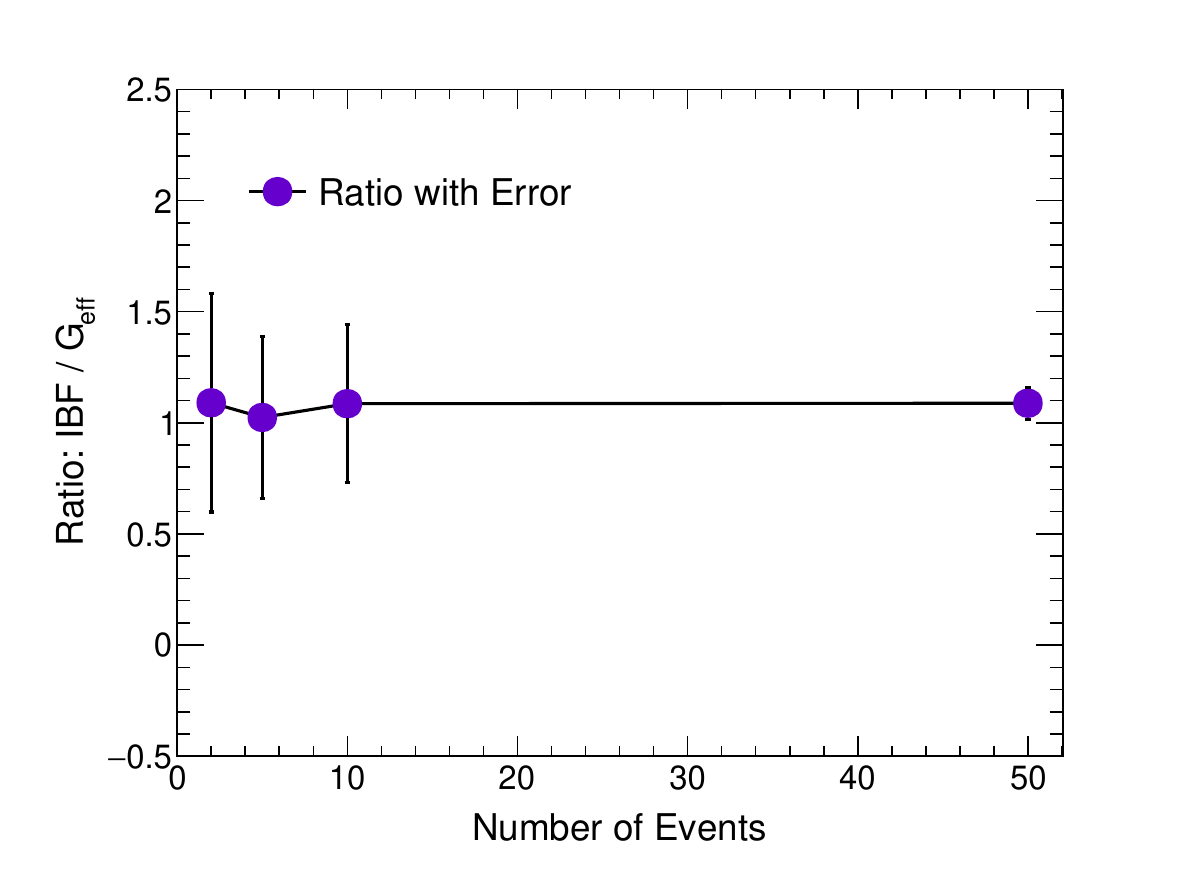}
  \caption{Ratio of ion backflow to effective gain for the standard Triple GEM configuration. The error bars represent statistical fluctuations, which decrease as the number of events increases.}
  \label{fig:ratio_BC}
\end{figure}

In continuation of the previous study, the ratio of ion backflow per event to the effective gain was calculated for the conventional triple-GEM detector with bi-conical hole geometry. The obtained ratio was close to unity. This result indicates a direct correlation between ion generation and charge amplification.
This behavior is anticipated in conventional triple GEM designs; however, it imposes constraints on applications that require reduced ion feedback into the drift area. Fig.~\ref{fig:ratio_BC} illustrates the statistical uncertainty in the ratio, which is large for a small number of events; however, the uncertainty decreases significantly as the number of events increases.
In this work, the ion backflow per event, normalized to the effective gain, is explicitly evaluated and adopted as a primary performance metric, enabling a quantitative, geometry-independent comparison among different detector configurations. This ratio provides a standardized measure of the ion feedback relation to the achieved charge amplification. A reduction in this metric indicates improved detector performance, reflecting the simultaneous increase in effective gain and reduction in ion transport into the drift region. Within this framework, a modified triple-GEM geometry employing single-conical holes is subsequently investigated. The ion-backflow-to-gain ratio is systematically analyzed for three lower copper layer thicknesses of 5~$\mu$m, 10~$\mu$m, and 15~$\mu$m. This comparative study evaluates the effectiveness of geometric modifications in reducing ion backflow while maintaining or enhancing the effective gain, thereby providing critical guidance for optimizing triple-GEM detectors for high-rate and precision experimental applications.

\subsection{Modified Triple GEM:
}
\begin{figure}[H]
  \centering
  \includegraphics[width=0.8\columnwidth]{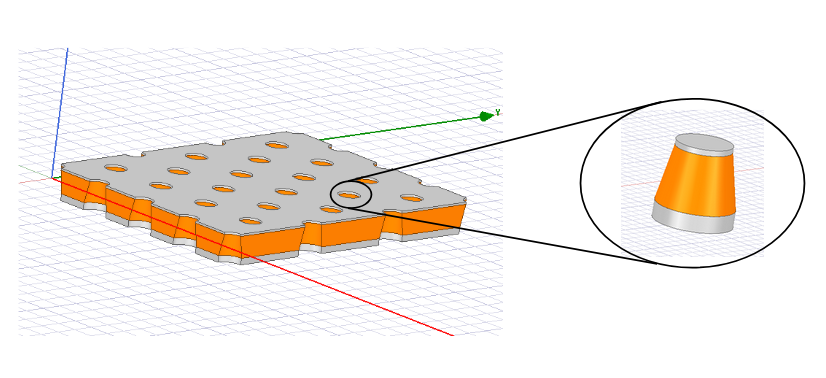}
  \caption{Schematic illustration of a GEM foil featuring a modified hole geometry, where the conventional bi-conical structure is replaced by a single-conical design. The lower copper layer thickness is systematically increased from the standard 5 $\mu$m to 10 $\mu$m and 15 $\mu$m in two successive steps.}
  \label{fig:SC_cell}
\end{figure}

The redesigned GEM foil shape was built and examined using ANSYS Maxwell. The modified hole shape and the respective foil structure are shown in Fig.~\ref{fig:SC_cell}.
Electric fields were established at the various detector locations within this framework. Furthermore, the electric-field distribution was kept identical for all three modified configurations, corresponding to lower copper thicknesses of 5 $\mu$m, 10 $\mu$m, and 15 $\mu$m. Maintaining the same electric field across these configurations ensures that any observed differences in detector performance arise solely from geometrical modifications rather than field variations, thereby enabling fair and physically meaningful comparisons between the different foil designs.

\begin{figure}[H]
  \centering
  \includegraphics[width=0.75\columnwidth]{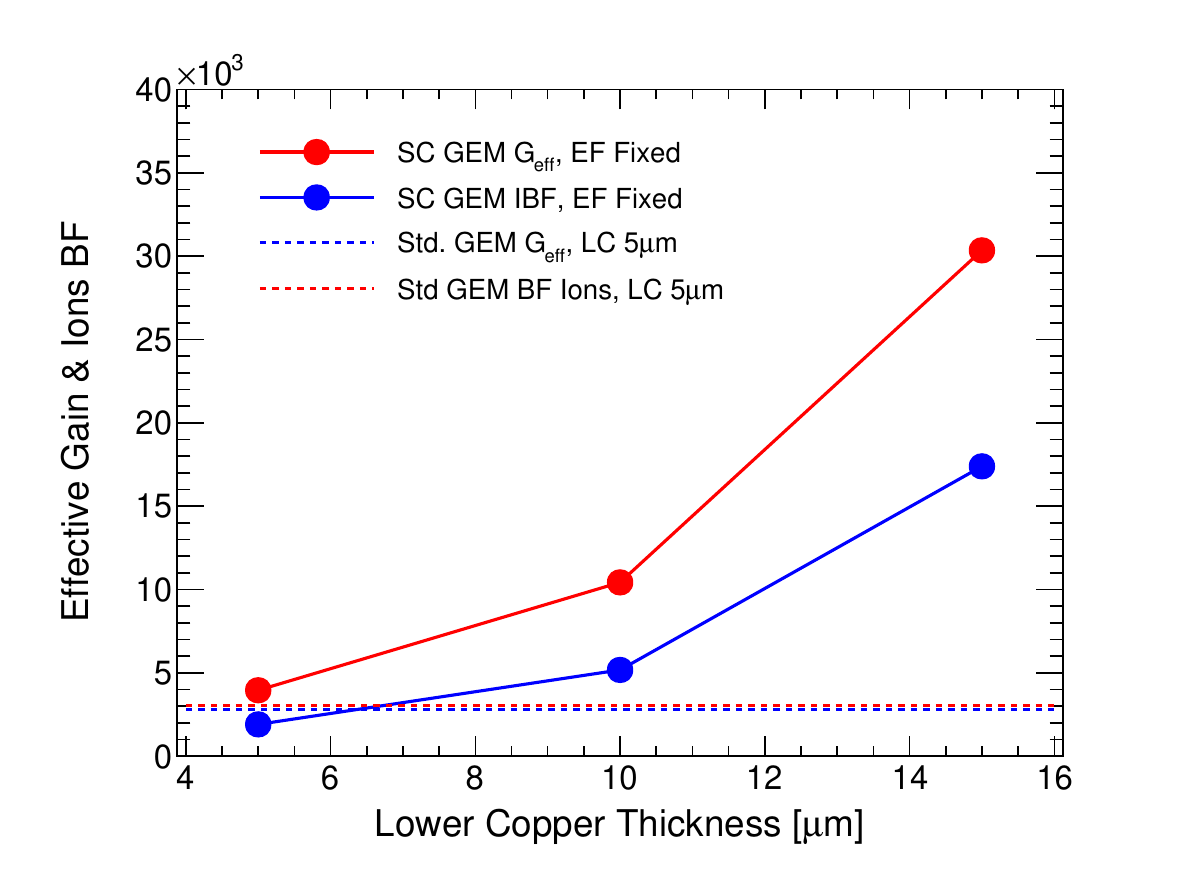}
  \caption{Trends of detector gain and ion backflow for single-conical and bi-conical triple-GEM configurations. Solid lines correspond to the modified single-conical triple-GEM, while dotted lines represent the standard bi-conical triple-GEM. The solid blue line shows the effective gain as a function of the lower copper thickness for the single-conical GEM, and the solid red line represents the corresponding ion backflow. The red dotted line indicates the ion backflow, and the blue dotted line the effective gain, for the bi-conical standard triple-GEM configuration.}
  \label{fig:sc_vs_dc}
\end{figure}

The detector gain and ion backflow were computed under these conditions. Subsequently, the mean electric field within all GEM foils and throughout all detector regions was assessed for this setup. To facilitate a uniform comparison across varying copper thicknesses, the electric field profiles acquired for the single-conical triple-GEM with a 5 $\mu$m lower copper layer were used as a basis for configurations with 10 $\mu$m and 15 $\mu$m lower copper thicknesses, with electrode voltages suitably modified in ANSYS Maxwell. This method ensured the preservation of uniform electric-field conditions across all configurations.
Upon establishing the requisite electric field distributions, accurate field maps were produced by recreating the revised GEM geometry in ANSYS Mechanical APDL. The field maps were later imported into Garfield$^{++}$, where comprehensive microscopic simulations were conducted. It includes the calculation of gain and ion backflow across various incident particle energies and simulated event statistics.
The detector gain was initially calculated as a function of the thickness of the lower copper layer of the GEM foil. The results indicated a distinct trend: as the lower copper thickness increased from 5 $\mu$m to 10 $\mu$m and 15 $\mu$m, the gain increased progressively. The increasing trend in detector gain is illustrated in Fig.~\ref{fig:sc_vs_dc}. A comparable methodology was utilized to examine the correlation between ion backflow (IBF) and increased copper thickness. The findings demonstrated that ion backflow escalated with greater copper thickness.

The reconfigured GEM foil shape featuring single-conical apertures was developed and examined. Initially, the identical electrode voltages used for the regular triple-GEM detector featuring bi-conical holes were applied to the single-conical triple-GEM configuration, which had both upper and lower copper thicknesses of 5 $\mu$m.

\begin{figure}[H]
  \centering
  \includegraphics[width=0.75\columnwidth]{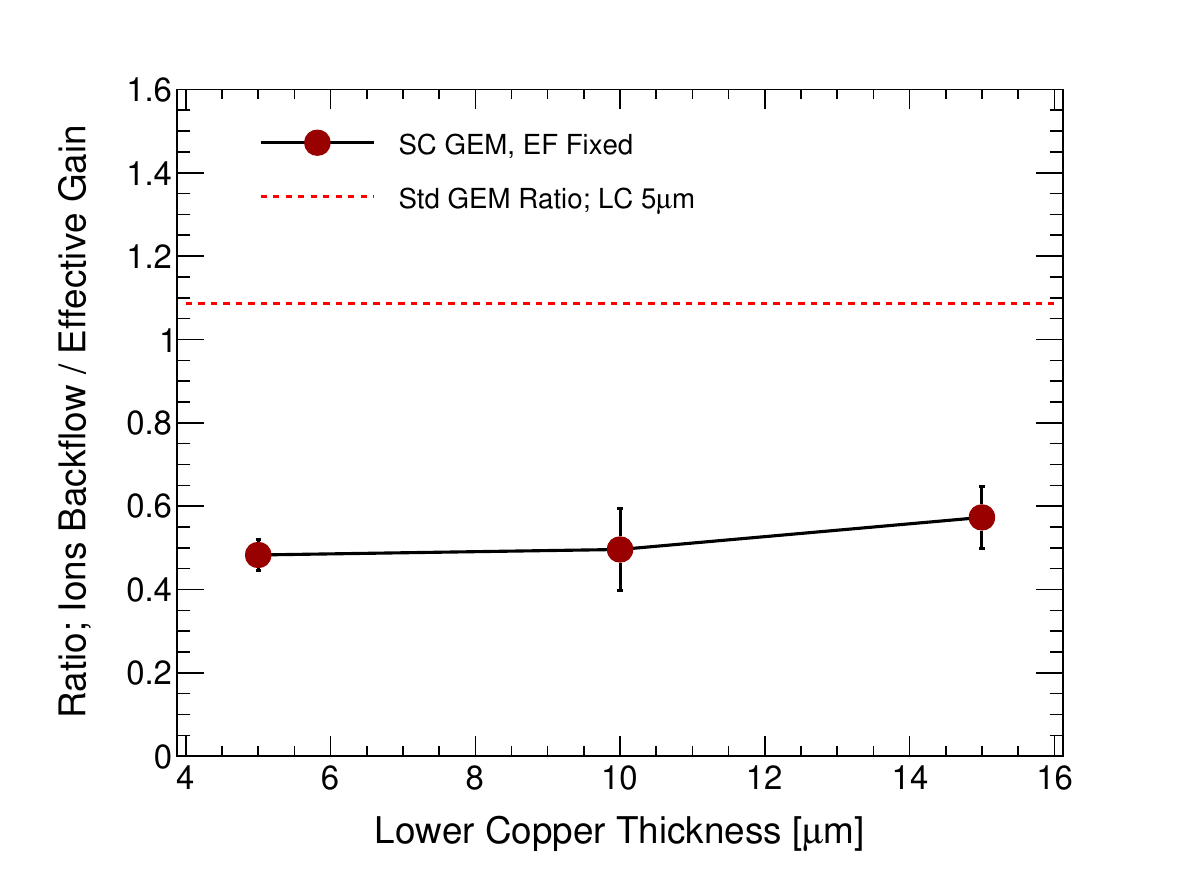}
  \vspace{-0.6cm}
  \caption{Ratio of ion backflow to effective gain for the modified Triple GEM as a function of lower copper thickness (blue line). The red-dotted horizontal line represents the corresponding value for a bi-conical Triple GEM. The plot provides a comparative assessment of the ion-backflow-to-gain ratio between the standard geometry and the newly proposed configuration.}
  \label{fig:ratio_sc}
\end{figure}

Nonetheless, a significant observation was that the elevation in gain rate was more pronounced than the associated rise in ion backflow. The primary figure of merit for GEM-based detectors, the ratio of ion backflow to gain, was calculated for each of the three thickness configurations to assess the detector's overall performance.  Across the various thicknesses, the ratio of ion backflow to gain remained between 0.50 and 0.55 as shown in Fig.~\ref{fig:ratio_sc}.  This ratio dropped by over 50\% in the new arrangement, according to a comparison between the original triple-GEM and the modified design.  This decrease, therefore, suggests a significant increase in overall efficiency of approximately 200\%.

 These findings conclusively demonstrate that the proposed improved GEM shape enhances the detector's overall gain while also effectively preventing ion backflow.  For high-rate investigations, where stable and accurate detector performance depends on sustaining high gain with little ion feedback, such advancements are crucial. Accordingly, the results indicate that this recently refined design presents a viable path for the upcoming generation of GEM-based detectors.
 \vspace{-0.9cm}
\subsection{Dependency of ion absorption on number of foils used:} In this section of the study, the total quantity of ions absorbed is analyzed as a function of the number of GEM foils utilized. The analysis evaluates the ion-absorption performance of a conventional bi-conical-hole GEM in comparison to a modified single-conical-hole GEM, in which the thickness of the lower copper layer has been increased.
\subsubsection{For bi-conical Single, Double, and Triple GEM}
Here, the total count of absorbed ions is assessed relative to the total number of ions generated during a specific event for the standard GEM utilizing the bi-conical hole structure. 

\begin{figure}[H]
  \centering
  \begin{subfigure}[b]{0.49\textwidth}
    \centering
    \includegraphics[width=\textwidth]{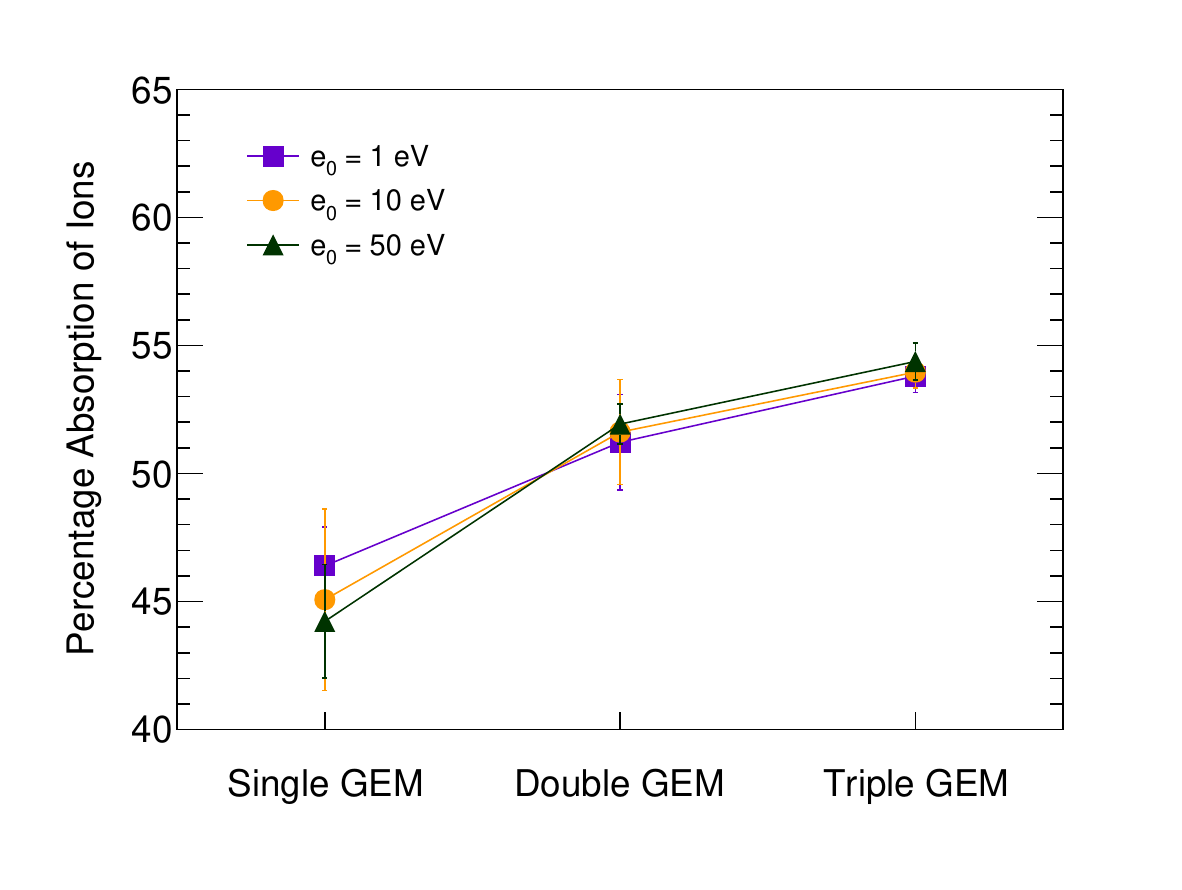}
  \caption{Percentage of ions absorbed as a function of the number of GEM foils is shown in the figure. A clear increasing trend is observed with an increasing number of foils. The three curves correspond to different initial electron energies: 1 eV (blue), 10 eV (orange), and 50 eV (green).}
  \label{fig:abs_foil}
  \end{subfigure}
  \hfill
  \begin{subfigure}[b]{0.49\textwidth}
    \centering
    \includegraphics[width=\textwidth]{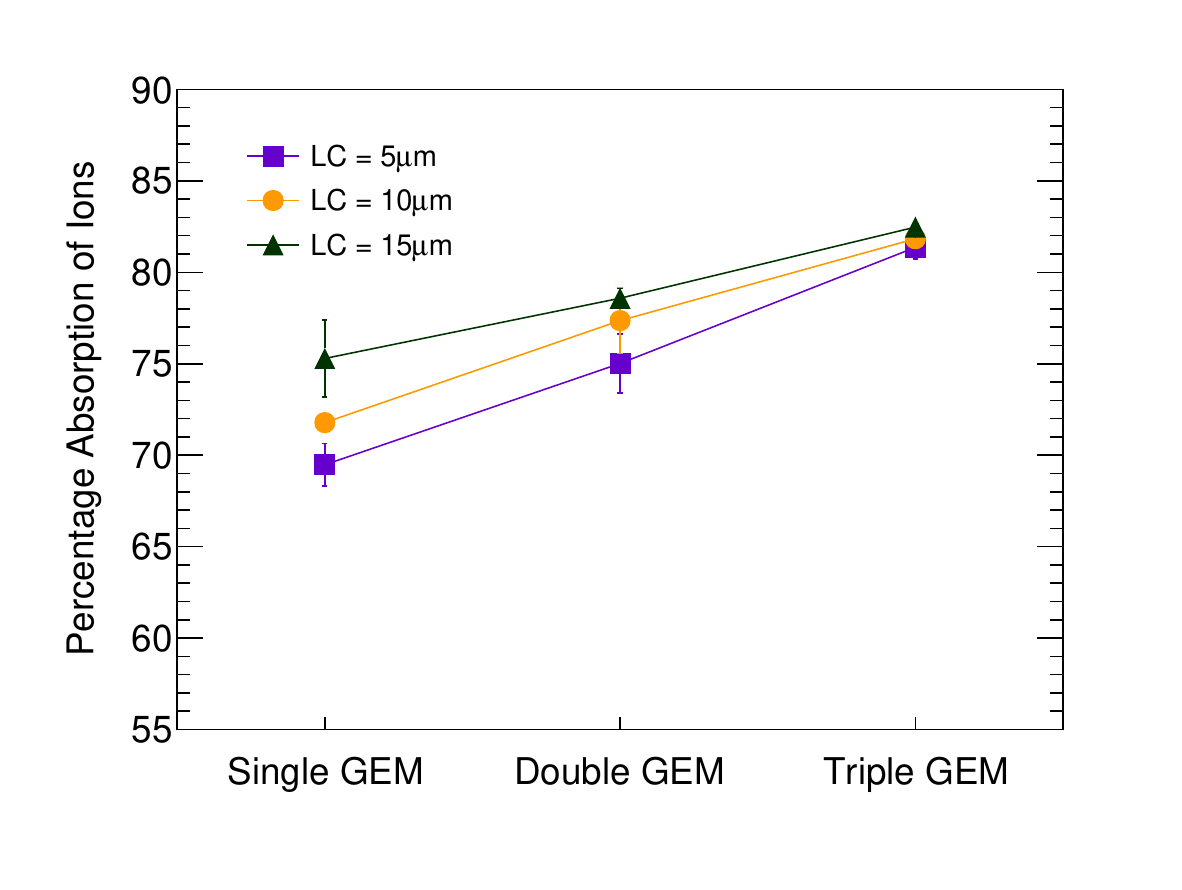}
  \caption{Ion absorption percentage as a function of the number of GEM foils for the modified single-conical GEM geometry. Results are shown for single-, double-, and triple-GEM configurations at a fixed primary electron energy for three different lower copper thicknesses: 5 $\mu$m, 10 $\mu$m, and 15 $\mu$m }
  \label{fig:IBF_SC}
  \end{subfigure}
  \caption{Ion absorption comparison}
  \label{fig:std_combined}
\end{figure}

Using this method, the percentage of ion absorption is determined for single-, double-, and triple-GEM configurations, demonstrating a distinct correlation between ion absorption and the number of GEM foils used. The identical analysis is conducted across three distinct primary electron energies, consistently reaffirming the observed phenomenon.
As demonstrated in Fig.~\ref{fig:abs_foil} , a notable enhancement in the ion absorption percentage is observed when transitioning from a single-GEM to a double-GEM configuration across all examined energies.

An additional increase in the ion absorption percentage is observed when transitioning from the double-GEM to the triple-GEM configuration; however, the extent of this increase is less pronounced than in the previous transition. Nevertheless, the overall trend clearly shows a consistent increase in ion absorption with the number of GEM foils.

These findings suggest that ion absorption is significantly influenced by the quantity of foils used. Increasing the quantity of GEM foils improves ion absorption, thereby reducing ion backflow.

\subsubsection{For Single conical Single, Double, and Triple GEM with variable lower copper thickness}
A comparable analysis is subsequently conducted for the modified GEM geometry incorporating single-conical perforations, examining single, double, and triple-GEM configurations. 

In this instance, the ion absorption percentage is assessed as a function of the number of foils at a constant primary electron energy, while the thickness of the lower copper layer is varied. Three lines are displayed, representing lower copper thicknesses of 5 $\mu$m, 10 $\mu$m, and 15 $\mu$m. In all three configurations, a general rise in ion absorption with increasing numbers of GEM foils is observed, consistent with expectations. Notably, the curve associated with a smaller copper thickness of 5 $\mu$m demonstrates the lowest ion absorption, whereas the maximum absorption is observed at a thickness of 15 $\mu$m, with the 10 $\mu$m case situated between these two values. This behavior is expected and indicates that increasing the lower copper thickness improves ion absorption efficacy.

When these results are compared to those obtained for the conventional bi-conical-hole GEM geometry, a comparable qualitative trend concerning the number of foils is observed. Nonetheless, the absolute ion absorption values in the modified single-conical geometry are markedly greater which can be seen in Fig.~\ref{fig:IBF_SC}.

This clearly demonstrates that the altered geometry offers enhanced suppression of ion reflux. Furthermore, this enhancement is accompanied by a significant increase in effective gain, demonstrating the superior overall performance of the modified GEM configuration.

\section{Summary and conclusions}
Extensive studies were conducted to investigate and optimize the geometrical parameters of triple-GEM detectors, with a particular emphasis on the effects of copper thickness and hole geometry on detector gain, ion backflow, and ion absorption. The performance of a standard triple-GEM detector employing a bi-conical hole structure was first examined, demonstrating stable gain and ion backflow behavior over a wide range of event statistics and incident primary electron energies. Both gain and ion backflow increased with increasing primary electron energy.

To further enhance detector performance, a modified triple-GEM geometry incorporating single-conical holes was developed and studied for three lower copper thicknesses: 5 $\mu$m, 10 $\mu$m, and 15 $\mu$m. For a meaningful comparison, the modified geometry was initially optimized using the ANSYS Maxwell to reproduce a comparable electric-field configuration. Subsequent Garfield$^{++}$ studies revealed that increasing the lower copper thickness leads to a systematic increase in effective gain, accompanied by a relatively modest increase in ion backflow.

In addition to gain and ion backflow, a detailed investigation of ion absorption was performed for both standard and modified geometries. The fraction of absorbed ions was measured for single-, double-, and triple-GEM configurations, clearly demonstrating that ion absorption increases with the number of GEM foils. This confirms that multi-foil configurations are more effective in ion trapping and backflow suppression. Notably, the modified single-conical geometry exhibits significantly higher ion absorption than the conventional bi-conical design, particularly at larger lower copper thicknesses, leading to a further reduction in ion backflow.

Overall, the combined analysis of gain, ion backflow, and ion absorption indicates that the modified single-conical GEM geometry outperforms the conventional bi-conical configuration. The ion-backflow-to-gain ratio is reduced by nearly a factor of two, relative to the standard design, which is expected to yield an improvement in detection efficiency of approximately 200\%. These results demonstrate that tailoring the hole geometry and copper thickness provides an effective strategy to enhance ion suppression while maintaining or improving detector gain. Consequently, the proposed triple-GEM configuration exhibits strong potential for high-rate particle detection applications that require improved accuracy, stability, and reduced ion feedback.

\section*{Acknowledgment}

The authors sincerely acknowledge the IIT Mandi HEP group for their insightful discussions and constructive feedback throughout this work. The author also extends sincere gratitude to the University Grants Commission–Council of Scientific and Industrial Research (UGC–CSIR) for their financial support through a Ph.D. fellowship. The authors also acknowledge IIT Mandi for providing access to computational resources and supercomputing facilities essential for this research.

\section*{Declaration of generative AI and AI-assisted technologies in the manuscript preparation process}

During the preparation of this work, the authors used QuillBot (Premium) for language refinement and grammar correction. After using this tool, the authors reviewed and edited the content as needed and take full responsibility for the content of the published article.

\bibliography{example}
\bibliographystyle{elsarticle-num}






\end{document}